\documentclass[twocolumn,prl,showpacs,amsmath,amssymb,nofootinbib]{revtex4}
\usepackage{amsthm,url,epsf}

\theoremstyle{plain}

\begin{document}
\title{Sealing Quantum Message By Quantum Code}
\author{H. F. Chau}
\email{hfchau@hkusua.hku.hk}
\affiliation{Department of Physics, University of Hong Kong, Pokfulam Road,
 Hong Kong.}
\date{\today}

\begin{abstract}
 Quantum error correcting code is a useful tool to combat noise in quantum
 computation. It is also an important ingredient in a number of unconditionally
 secure quantum key distribution schemes. Here, I am going to show that quantum
 code can also be used to seal a quantum message. Specifically, every one can
 still read the content of the sealed quantum message. But, any such attempt
 can be detected by an authorized verifier with an exponentially close to one
 probability.
\end{abstract}

\maketitle
\emph{Introduction ---}
 We sometimes put an important document, such as a will, in an envelop sealed
 with molten wax so that others can open it only by breaking the wax. The
 sealed envelop, therefore, acts like a witness of whether the document has
 been open.

 Clearly, it is meaningful and useful to extend the concept of physical wax
 seal to the digital world. Yet, no digital sealing scheme is unconditionally
 secure in the classical digital world as one can, in principle, copy all the
 bits without being caught.

 Recently, Bechmann-Pasquinucci examined the possibility of sealing a classical
 digital signal using quantum mechanics. Specifically, he proposed a way to
 represent one bit of classical signal by three qubits out of which one of them
 is erroneous. Using single qubit measurement along the standard basis plus the
 classical $[3,1,3]_2$ majority vote code, everyone can obtain the original
 classical bit with certainty. And at the same time, an authorized verifier,
 who knows some extra information on the erroneous qubit, is able to check if
 someone has extracted the encoded classical bit with non-negligible
 probability \cite{quantseal}.

 Nevertheless, Bechmann-Pasquinucci's scheme is far from a good quantum seal as
 there is a non-negligible chance of knowing the original bit without being
 caught.\footnote{Actually, Bechmann-Pasquinucci was more interested in the
 security of obtaining the entire classical bit string by which each bit is
 encoded using his scheme. However, this security requirement is too relaxed as
 the ability to know just a single bit of an important document without being
 caught may already have serious sequences.} For instance, one may randomly
 pick one of the qubits and measure its state along the standard basis. In this
 way, the chance of knowing the original classical bit correctly without being
 caught equals $2/3 + (1/3) (1/2) (1/2) = 0.75$. Furthermore, it is not clear
 how to seal quantum information in Bechmann-Pasquinucci's scheme.

 In this Letter, I report a quantum seal scheme using quantum error correcting
 code. This scheme can seal both classical and quantum signals. I also prove
 that any person other than an authorized verifier has an exponentially small
 chance of knowing the original signal without being caught. This remains true
 even when the person has unlimited computational power.

\begin{figure}[t]
 \epsfxsize=8cm
 \epsfbox{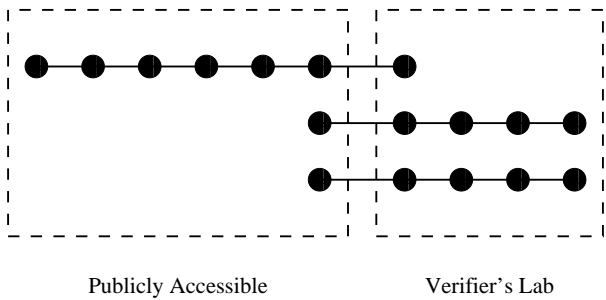}
 \caption{As an example, I schematically represent the quantum seal using
  the $[[7,1,3]]_2$ CSS code and the $[[5,1,3]]_2$ perfect code in this figure.
  Each dot represents a qubit and the solid line joining the dots represents
  the entangled qubits in $[[7,1,3]]_2$ or $[5,1,3]]_2$ code. Qubits in the
  left dashed box are publicly accessible while those in the right dashed box
  are available only to an authorized verifier.}
 \label{F:Seal}
\end{figure}

\emph{The Quantum Seal ---}
 Let $|\psi\rangle$ be the state of a qubit that Alice wants to make public so
 that anyone may use it. (Note that because of the no cloning theorem, Alice
 may know nothing about the state $|\psi\rangle$.) To seal the state, Alice
 publicly announces a $[[n,1,d]]_2$ Calderbank-Shor-Steane (CSS) code
 \cite{cs,steane1,steane2} with $d\geq 0.11n$ and uses it to encode $|\psi
 \rangle$. (The existence of such a code is guaranteed by a theorem of
 Gottesman in Ref.~\cite{qhammingbound}.) The encoded state is denoted by
 $|\psi\rangle_\textrm{\scriptsize L}$. Alice further chooses a $[[n',1,3]]_2$
 stabilizer code and uses it to encode $t \equiv \left\lfloor (d-1)/2
 \right\rfloor$ copies of $|0\rangle$. Such an encoded state is denoted by $|0
 \rangle_{\textrm{\scriptsize L}'}^{\otimes t}$. As depicted in
 Fig.~\ref{F:Seal}, Alice randomly selects $(n-t)$ qubits from those used to
 represent $|\psi\rangle_\textrm{\scriptsize L}$ and $t$ qubits each from a
 separate copy of $|0\rangle_{\textrm{\scriptsize L}'}$. She makes these $n$
 selected qubits publicly accessible. At the same time, she keeps all the
 remaining $n' t$ qubits (which consist of the $t$ qubits used to encoded
 $|\psi\rangle_\textrm{\scriptsize L}$ and the remaining $(n' -1) t$ qubits
 used to encode $t$ copies of $|0\rangle$) in a secure place. This marks the
 end of the quantum sealing phase. Clearly, the security of the quantum seal
 originates from the secrecy of which of the $(n-t)$ publicly available qubits
 are used to encode $|\psi\rangle$.

 To open the seal and obtain the original quantum state, Bob simply applies the
 standard quantum error correction procedure for the $[[n,1,d]]_2$ CSS code to
 the $n$ publicly accessible qubits. This method works because there are only
 $t$ out of the $n$ qubits at unknown locations that are not used to encode
 $|\psi\rangle$.

 To check if the seal is opened, we need an authorized verifier who has access
 to the remaining $n' t$ qubits and knows the locations of the $t$ publicly
 accessible qubits that are not used to encode $|\psi\rangle$. Clearly, if no
 one has touched the $n$ publicly available qubits, the $(n+n' t)$ qubits
 accessible by the verifier should be in the state $|\psi
 \rangle_\textrm{\scriptsize L} \otimes |0
 \rangle_{\textrm{\scriptsize L}'}^{\otimes t}$. Therefore, the verifier
 accepts the seal as unbroken only if the $[[n,1,d]]_2$ and the $[[n',1,3]]_2$
 stabilizer code error syndrome measurements reveal that all the $(n+n' t)$
 qubits are error-free.
 
\emph{Security Of The Quantum Seal ---}
 Now, I am going to show that the above quantum sealing scheme is
 unconditionally secure. Specifically, for any security parameters
 $(\epsilon_p,\epsilon_I)$ , there exists a quantum sealing scheme using a
 sufficiently long quantum codeword length $n$ such that whenever Bob (who is
 not an verifier) applies a cheating strategy whose probability of success is
 at least $\epsilon_p$, his information on the state $|\psi\rangle$ is less
 than $\epsilon_I$. Moreover, this is true even if he has unlimited
 computational power.

 I prove the unconditional security of the above scheme by reduction. First, I
 consider a revised quantum sealing scheme and show that it is as secure as the
 original one introduced above. The encoding procedure of the revised scheme is
 the same as the original one. But in the verification procedure of the revised
 scheme, the verifier also measures the encoded spin flip operator for the
 $[[n,1,d]]_2$ and the $[[n',1,3]]_2$ stabilizer codes in addition to the error
 syndrome measurement. Yet, the verifier accepts the quantum seal as unbroken
 based only on the result of the error syndrome measurement in exactly the same
 way as in the original scheme. Note that the encoded spin flip operations
 commute with the stabilizer and the acceptance criterion of the revised scheme
 does not depend on the result of the encoded spin flip measurement.
 Consequently, any cheating strategy will have an equal chance to pass through
 the verification test and will reveal an equal amount of information on the
 sealed quantum state when applied to the original and the revised schemes. In
 this respect, the two schemes are equally secure; and it suffices to prove the
 security of the revised scheme.
 
 Second, I reduce every quantum cheating strategy for the revised scheme to a
 corresponding classical cheating strategy. Observe that there are $(n+n' t)$
 independent stabilizers and encoded spin flip operations, all the $(n+n' t)$
 qubits are measured in the revised scheme during the verification phase.
 Hence, following the argument of Lo and Chau in Ref.~\cite{lochausci}, any
 quantum cheating strategy for the revised scheme can be reduced to a classical
 probabilistic one. More precisely, all cheating strategies can be generated by
 the classical deterministic strategies $\bigotimes_{i=1}^n s_i$ acting on the
 $n$ qubits accessible by Bob, where $s_i \in \{ I, \sigma_x, \sigma_y,
 \sigma_z \}$.

 Third, I analyze what type of classical cheating strategy reveals the
 information on the locations of the $t$ publicly accessible qubits that are
 not used to encode the sealed state $|\psi\rangle$. As the $t$ qubits taken
 from $|0\rangle_{\textrm{\scriptsize L}'}^{\otimes t}$ are unentangled with
 the $(n-t)$ qubits taken from $|\psi\rangle_\textrm{\scriptsize L}$, the
 entropy and hence the information obtained from the measurement of those
 qubits coming from $|\psi\rangle_\textrm{\scriptsize L}$ and $|0
 \rangle_{\textrm{\scriptsize L}'}^{\otimes t}$ using any classical strategy
 are independent. Since the code $[[n,1,d]]_2$ used is publicly known and all
 the qubits used to encode $|\psi\rangle$ do not suffer from any quantum error,
 any measurement of the $(n-t)$ publicly available qubits taken from $|\psi
 \rangle_\textrm{\scriptsize L}$ only reveals information on the sealed quantum
 state $|\psi\rangle$. Similarly, any measurement of the $t$ qubits taken from
 $|0\rangle_{\textrm{\scriptsize L}'}^{\otimes t}$ only reveals information on
 the location and the nature of the ``quantum error'' by regarding the $n$
 publicly available qubits as a $[[n,1,d]]_2$ codeword.

 Fourth, I show that the verification test put a stringent limit on the
 information obtained by any classical cheating strategy. Recall that only one
 qubit for each $[[n',1,3]]_2$ encoded $|0\rangle$ are publicly accessible.
 Consequently, amongst the deterministic cheating strategies
 $\bigotimes_{i=1}^n s_i$, only those that cause no error in all the $t$ qubits
 taken from $|0\rangle_{\textrm{\scriptsize L}'}^{\otimes t}$ can pass the
 verification test. So, in order to have at least $\epsilon_p$ chance of
 passing the verification test, Bob cannot put too much weight on those
 cheating strategies that cause error in these $t$ qubits. Similarly, only
 those errors in the $(n-t)$ qubits taken from $|\psi
 \rangle_\textrm{\scriptsize L}$ that commute with the stabilizer of
 $[[n,1,d]]_2$ can pass the test. Recall that operations commuting with the
 stabilizer of $[[n,1,d]]_2$ are generated by the stabilizer and the encoded
 operations. Besides, an error in the stabilizer of a code simply permutes the
 stabilizer itself \cite{gott_thesis}. Therefore, effectively the types of
 action on $|\psi\rangle_\textrm{\scriptsize L}$ that can pass the verification
 test are those generated by the encoded operations of $[[n,1,d]]_2$.

 By the Holevo theorem \cite{holevobound}, Bob's information $I_e$ on the
 locations of the $t$ publicly accessible qubits that are not used to encode
 $|\psi\rangle$ is upper bounded by the entropy of the reduced density matrix
 of the $n' t$ qubits originally used to encode $|0\rangle^{\otimes t}$.
 Therefore, $I_e$ is upper bounded by the entropy of the reduced density matrix
 $\mbox{diag} [a,(1-a)/(2^t-1),(1-a)/(2^t-1),\ldots ,(1-a)/(2^t-1)]$. In other
 words,
\begin{eqnarray}
 I_e & \leq & -a \log_2 a - (1-a) \log_2 [(1-a)/(2^t - 1)] \nonumber \\
 & < & H(a) + t(1-a) , \label{E:inequality1}
\end{eqnarray}
 where $H(a) \equiv -a \log_2 a - (1-a) \log_2 (1-a)$ and $a$ is the
 probability of choosing a deterministic strategy that causes no error in all
 the $t$ qubits taken from $|0\rangle_{\textrm{\scriptsize L}'}^{\otimes t}$.

 If the strategy passes the verification test with probability at least
 $\epsilon_p$, we demand $\epsilon_p \leq a \leq 1$. Provided that Alice
 chooses a sufficiently long code $[[n,1,d]]_2$ in such a way that $t \equiv
 \left\lfloor (d-1)/2 \right\rfloor$ satisfies
\begin{equation}
 \epsilon_p > 1 / (1+2^t) , \label{E:condition}
\end{equation}
 Bob's information on the locations of those $t$ qubits in
 Eq.~(\ref{E:inequality1}) is upper bounded by
\begin{equation}
 I_e < H(\epsilon_p) + (1-\epsilon_p) t \equiv I_\textrm{\scriptsize bound}
 (\epsilon_p) . \label{E:inequality2}
\end{equation} 
 Consequently, for any cheating strategy that passes the verification test with
 probability at least $\epsilon_p$, the locations of at most
 $I_\textrm{\scriptsize bound} (\epsilon_p)$ of out the $t$ publicly accessible
 qubits taken from $|0\rangle_{\textrm{\scriptsize L}'}^{\otimes t}$ are known
 to Bob.

 Finally, I am ready to bound the amount of information on the sealed quantum
 state revealed by a cheating strategy that passes through the verification
 test with probability at least $\epsilon_p$. But before I do so, let me
 summarize the situation after Bob passes the verification test. Bob knows the
 locations of at most $I_\textrm{\scriptsize bound} (\epsilon_p)$ out of the
 $n$ publicly accessible qubits that are not used to encode $|\psi\rangle$.
 Denote $\rho$ the covering radius of the $[[n,1,d]]_2$ quantum code. That is
 to say, $\rho$ is the smallest integer such that ${\mathbb F}_4^n$ equals the
 union of spheres of radius $\rho$ centered at the vectors spanned by the
 stabilizers and encoded operations of the $[[n,1,d]]_2$ CSS code over
 ${\mathbb F}_4$. In simple terms, $\rho$ is nothing but the maximum number of
 errors the quantum code $[[n,1,d]]_2$ can handle. Thus, in order to recover
 the sealed quantum state $|\psi\rangle$, Bob has to determine the locations of
 at least $(n-\rho)$ out of the remaining $[n-I_\textrm{\scriptsize bound}
 (\epsilon_p)]$ qubits that are used to encode $|\psi\rangle$. More
 importantly, he has to do so in the presence of at least $[t-
 I_\textrm{\scriptsize bound} (\epsilon_p)]$ qubits chosen from $|0
 \rangle_{\textrm{\scriptsize L}'}^{\otimes t}$ without affecting them.

 Since the locations of these $[t-I_\textrm{\scriptsize bound} (\epsilon_p)]$
 qubits are unknown to Bob, this problem is equivalent to picking $(n-\rho)$
 good balls out of an urn of $[n-I_\textrm{\scriptsize bound} (\epsilon_p)]$
 balls with $[t-I_\textrm{\scriptsize bound} (\epsilon_p)]$ of them being bad.
 The probability of correctly picking the good balls is given by
\begin{eqnarray}
 p & = & \prod_{i=0}^{n-\rho-1} \frac{(n-t-i)}{[n-I_\textrm{\tiny bound}
  (\epsilon_p)-i]} \nonumber \\
 & \leq & \left[ \frac{n-t}{n-I_\textrm{\tiny bound} (\epsilon_p)} \right]^{n-
  \rho} . \label{E:bound_prob}
\end{eqnarray}

 The $[[n,1,d]]_2$ CSS code can be regarded as a classical $(n,\left\lceil
 [n+1]/2 \right\rceil)$ code over ${\mathbb F}_4$ \cite{steane2,gott_thesis}.
 Applying the redundancy bound \cite{redundancy_bound} to this classical code,
 I conclude that
\begin{equation}
 \rho \leq n - \left\lceil \frac{n+1}{2} \right\rceil = \left\lfloor \frac{n-
 1}{2} \right\rfloor . \label{E:redundany_bound}
\end{equation}

 Recall that Bob may obtain some information on $|\psi\rangle$ only when he
 makes use of at least $(n-\rho)$ out of the $n$ publicly accessible qubits.
 Therefore, from Eqs.~(\ref{E:bound_prob}) and~(\ref{E:redundany_bound}), Bob's
 information $I_{|\psi\rangle}$ on $|\psi\rangle$ is upper bounded by 
\begin{equation}
 I_{|\psi\rangle} \leq \left[ \frac{n-t}{n-I_\textrm{\tiny bound} (\epsilon_p)}
 \right]^{\left\lceil (n+1)/2 \right\rceil} . \label{E:I_I}
\end{equation}
 Remember that one can always choose $t/n$ to be at least $0.5 H^{-1} (1/2)
 \approx 0.055$ \cite{qhammingbound}. Besides, Eq.~(\ref{E:inequality2})
 implies that $I_\textrm{\scriptsize bound} (\epsilon_p)$ increases linearly
 with $t$. Since $I_\textrm{\scriptsize bound} (\epsilon_p) < t$ whenever
 Eq.~(\ref{E:condition}) holds, I conclude that $(n-t)/[n-
 I_\textrm{\scriptsize bound} (\epsilon_p)]$ is upper bounded by a positive
 number $\alpha (\epsilon_p) < 1$ for sufficiently large $n$. More importantly,
 $\alpha (\epsilon_p)$ is independent of $n$. (Actually, $\alpha (0) \leq (1-
 0.055)/(1+0.055) < 0.896$.) Therefore, by choosing
\begin{equation}
 n \geq 2 \log_{\alpha (\epsilon_p)} \left( \frac{1}{\epsilon_I} \right) ,
 \label{E:bound_n}
\end{equation}
 Bob's information on the sealed quantum state $|\psi\rangle$ is less than
 $\epsilon_I$.
 
 In conclusion, I show that for any security parameters $(\epsilon_p,
 \epsilon_I)$, there is an unconditionally secure quantum seal with $n$
 satisfying Eqs.~(\ref{E:condition}) and~(\ref{E:bound_n}) such that whenever
 Bob uses a cheating strategy that passes with probability at least
 $\epsilon_p$, his information on the sealed quantum state $|\psi\rangle$ is
 less than $\epsilon_I$.

\emph{Discussions ---}
 Three remarks are in place. First, although I focus the discussion on sealing
 a pure quantum state, the analysis above applies equally well to mixed state.
 Hence, the quantum sealing scheme reported here is also valid for sealing
 mixed state.

 Second, since the verification test is nothing but an error syndrome
 measurement procedure, the quantum sealing scheme works equally well if the
 encoded state $|0\rangle_{\textrm{\scriptsize L}'}^{\otimes t}$ is replaced by
 $\bigotimes_{i=1}^t |\phi_i\rangle_{\textrm{\scriptsize L}'}$ for some pure
 states $|\phi_i\rangle$ ($i=1,2,\ldots ,t$).

 Third, all versions of quantum sealing schemes introduced in the Letter uses
 two stabilizer codes. It is more efficient to seal a quantum state by using
 just one stabilizer code. That is to say, one encodes the quantum state to be
 sealed by a $[[n,1,d]]_2$ code and randomly replace $t\equiv \left\lfloor
 (d-1)/2 \right\rfloor$ qubits in the codeword by some randomly chosen pure
 states similar to the Bechmann-Pasquinucci scheme \cite{quantseal}. It is
 instructive to investigate the unconditional security of this scheme by
 suitably bounding the information on the locations of the $t$ replaced qubits.


\bibliography{qc32.1}
\end{document}